\documentclass[11pt]{article}%
\usepackage{amsmath,amssymb,color,graphicx,sectsty}
\usepackage{geometry}
\newcommand{\be}{\begin{equation}}
\newcommand{\bea}{\begin{eqnarray}}
\newcommand{\ee}{\end{equation}}
\newcommand{\eea}{\end{eqnarray}}

\newcommand{\del}{\delta}

\newcommand{\calC}{\mathcal{C}}

\newcommand{\calS}{\mathcal{S}}

\newcommand{\dr}{\,\text{d}r}

\newcommand{\dtheta}{\,\text{d}\theta}

\newcommand{\drdtheta}{\,\text{d}r\mskip1mu\text{d}\theta}
\newcommand{\drhodtheta}{\,\text{d}\rho\mskip1.75mu\text{d}\theta}

\newcommand{\lint}{\int_0^{2\pi}}

\newcommand{\aintdim}{\int_0^{2\pi}\mskip-8mu\int_0^1}
\newcommand{\crig}{c\mskip1.5mu}

\newcommand{\go}{\mskip1.0mu}

\newcommand{\half}{{\textstyle{\frac{1}{2}}}}

\newcommand{\abs}[1]{|#1\rvert}

\newcommand{\bfb}{\boldsymbol{b}}

\newcommand{\bfd}{\boldsymbol{d}}
\newcommand{\bfe}{\boldsymbol{e}}

\newcommand{\bft}{\boldsymbol{t}}
\newcommand{\bfx}{\boldsymbol{x}}

\newcommand{\bfm}{\boldsymbol{m}}
\newcommand{\bfn}{\boldsymbol{n}}

\newcommand{\bfp}{\boldsymbol{p}}

\newcommand{\bfu}{\boldsymbol{u}}

\newcommand{\bfw}{\boldsymbol{w}}

\newcommand{\bfnu}{\boldsymbol{\nu}}

\newcommand{\bfeta}{\boldsymbol{\eta}}

\newcommand{\bfxi}{\boldsymbol{\xi}}
\newcommand{\bfz}{\boldsymbol{\zeta}}

\newcommand{\xir}{\bfxi_r}
\newcommand{\xirr}{\bfxi_{rr}}

\newcommand{\xit}{\bfxi_{\theta}}
\newcommand{\xitt}{\bfxi_{\theta \theta}}
\newcommand{\xittt}{\bfxi_{\theta \theta \theta}}
\newcommand{\xitttt}{\bfxi_{\theta \theta \theta \theta}}

\newcommand{\xirt}{\bfxi_{r\theta}}

\newcommand{\reqone}{\scriptstyle r\mskip1.75mu=\mskip0.75mu1}

\newcommand{\vit}{v_{\theta}}
\newcommand{\vitt}{v_{\theta \theta}}
\newcommand{\vittt}{v_{\theta \theta  \theta}}
\newcommand{\vitttt}{v_{\theta \theta  \theta \theta}}

\newcommand{\wir}{w_r}
\newcommand{\wirr}{w_{rr}}
\newcommand{\wit}{w_{\theta}}
\newcommand{\witt}{w_{\theta \theta}}
\newcommand{\wittt}{w_{\theta \theta  \theta}}
\newcommand{\witttt}{w_{\theta \theta  \theta \theta}}

\newcommand{\uit}{u_{\theta}}

\newcommand{\uro}{\bfeta_r}
\newcommand{\urro}{\bfeta_{rr}}
\newcommand{\ut}{\bfeta_{\theta}}
\newcommand{\utt}{\bfeta_{\theta\theta}}
\newcommand{\uttt}{\bfeta_{\theta\theta\theta}}
\newcommand{\utttt}{\bfeta_{\theta\theta\theta \theta}}

\newcommand{\wt}{\bfw_{\theta}}
\newcommand{\wtt}{\bfw_{\theta \theta}}
\newcommand{\wttt}{\bfw_{\theta \theta \theta}}

\newfont{\tenbss}{bbmss12}

\newfont{\tenbfsl}{cmbxti12}

\newcommand{\ctimes}{\mskip-1.25mu\times\mskip-1.25mu}

\newcommand{\sperp}{\scriptscriptstyle\perp}
\newcommand{\er}{\bfe}
\newcommand{\et}{\bfe^{\sperp}}
\newcommand{\ez}{\bfe \times \bfe^{\sperp}}

\begin{document}

\title{Euler--Plateau, with a twist}

\author{A. Biria$^{*}$ and E. Fried$^{**}$
\\[2pt]
\small$^{*}$\emph{Department of Mechanical Engineering}
\\[-2pt]
\small\emph{McGill University}
\\[-2pt]\small
\emph{Montr\'eal}, \emph{Qu\'ebec}, \emph{Canada H3A 2K6}
\\[2pt]
\small$^{**}$\emph{Mathematical Soft Matter Unit}
\\[-2pt]\small\emph{Okinawa Institute of Science and Technology} 
\\[-2pt]
\small\emph{Okinawa}, \emph{Japan 904-0495}}

\date{}

\maketitle

\begin{abstract}
\noindent
A generalization of the Euler--Plateau problem to account for the energy contribution due to twisting of the bounding loop is proposed. Euler--Lagrange equations are derived in a parameterized setting and a bifurcation analysis is performed. A pair of dimensionless parameters govern bifurcations from a flat, circular ground state. While one of these is familiar from the Euler--Plateau problem, the other encompasses information about the ratio of the twisting to bending rigidities, twist, and size of the loop. For sufficiently small values of the latter parameter, two separate groups of bifurcation modes are identified. On the other hand, for values greater than the critical twist arising in Michell's problem of the bifurcation of a twisted annular ring, only one bifurcation mode exists. Bifurcation diagrams indicate that a loop with greater twisting rigidity shows more resistance to transverse buckling. However, a twisted and closed filament spanned by a surface  endowed with uniform surface tension buckles at a twist less than the critical value for an elastic ring. 
\end{abstract}

\section{Introduction}
Plateau~\cite{p} observed that a rigid wire frame, regardless of its shape, is spanned by at least one soap film. In the famous Plateau problem, the frame is modeled as a  given closed, rectifiable, Jordan curve and the soap film is modeled as a surface of zero mean-curvature. A century after Plateau's observation, Alt~\cite{Alt} posed a variation of the Plateau problem, known as the ``thread problem,'' in which one or more segments of the supporting frame are replaced by twist-free, inextensible threads of prescribed lengths, with shapes to be determined. The thread problem has interested mathematicians for several decades, leading to various investigations of the existence and regularity of solutions, as detailed by Dierkes, Hildebrandt \& Sauvigny~\cite{DH1} and Dierkes, Hildebrandt \& Tromba~\cite{DH2}. A specialization of the thread problem in which the frame contains no rigid segments was first considered by Bernatzki and Ye~\cite{BY}. In this problem, the flexural resistance of the bounding loop competes with surface tension, which may deform the boundary to reduce the area of the spanning surface. Recognizing that it couples aspects of Euler's elastica with Plateau's problem, Giomi \& Mahadevan~\cite{gm} referred to this specialization of the thread problem as the ``Euler--Plateau problem.'' Issues related to the stability and bifurcation of flat, circular solutions of the Euler--Plateau problem were considered by Giomi \& Mahadevan~\cite{gm} and, subsequently, by Chen \& Fried~\cite{CF}. Experimental realizations of the problem, in which loops of fishing line were dipped into and extracted from soapy water, {were} considered by Giomi \& Mahadevan~\cite{gm} and Mora et. al.~\cite{MF}.

The experiments and numerical simulations of Giomi \& Mahadevan~\cite{gm} predict a rich spectrum of equilibrium configurations, including non-planar saddle-like surfaces and twisted figure-eights. The observation of non-planar and twisted conformations immediately raises the question of what happens if the bounding loop resists twist in addition to bending? Also of interest is the related question of what happens if the fishing line is twisted before its ends are joined to form a loop. 
These questions suggest a further generalization of the Euler--Plateau problem wherein the role of twist is taken into account. In that this involves issues similar to those encountered in Michell's~\cite{Michell} studies of the bifurcation of a twisted annular ring (see also the later works of Zajac \cite{Zajac} and Benham \cite{Benham}), it seems reasonable to {refer} to the relevant generalization as the `Euler--Plateau--Michell problem.' 

Applications involving {biofilaments} (see, for example, Schlick \cite{Schlick} Goriely \& Tabor \cite{GT}, Coleman \& Swigon \cite{CS}, Hoffman, Manning \& Maddocks~\cite{hmm}, and Thompson, van der Heijden \& Neukirch~\cite{thn}), have recently inspired extensive interest in twisted elastic rings. In the other hand, soap films have long served as prototypes for material interfaces and fluid membranes such as lipid bilayers. {A linkage between the two therefore} provides a canonical model for various physical and biological systems in which surface and boundary energies compete. Prominent examples of such systems include discoidal high-density lipoprotein particles (Catte et al.~\cite{Catte}), which consist of lipid bilayers bound by chiral protein belts,\footnote{Prior to the introduction of the Euler--Plateau problem by Giomi \& Mahadevan~\cite{gm}, Catte et al.~\cite{Catte} proposed that observed buckling transitions between flat, circular and saddle-like configurations of discoidal high-density lipoprotein particles might be aptly modeled by treating the lipid bilayer as a soap film and the protein belt as a twist-free, inextensible rod. See Maleki \& Fried~\cite{mf}, who study an augmentation of the Euler--Plateau problem that accounts for the resistance of the bilayer to bending and the resistance of the protein belt to kinking.} and bacterial biofilms (Whitchurch et al.~\cite{biofilm}), which combine hydrated matrices of microorganisms and DNA. 
Another example is the convoluted intestinal canal anchored by the dorsal mesentery membrane. Savin et. al~\cite{gut} discovered that the force exerted by the attached mesentery is responsible for the chirality and looping of vertebrate intestine.    

The remainder of this paper is organized as follows. The basic assumptions regarding the models for the bounding loop and spanning surface appear in Section~\ref{prelim}, along with some essential geometric preliminaries. For an unshearable and isotropic rod with uniform circular cross-section, it suffices to consider a single director in the cross section of the bounding loop or, equivalently, the angle which such a director forms with the Frenet--Serret frame of the centerline of that loop. A model net potential-energy is provided in Section~\ref{generalformulation}. Following Giomi \& Mahadevan~\cite{gm}, who disregard intrinsic curvature, we consider neither intrinsic curvature nor intrinsic twist. Two dimensionless parameters in addition to that arising in the Euler--Plateau problem are identified. While the first of these accounts for the ratio between the twisting and bending rigidities of the bounding loop, the second combines the first with the dimensionless twist. A parametric version of the problem, is presented in Section~\ref{generalformulation}. In this setting, which draws on the works of Alt~\cite{Alt}, Hildebrandt~\cite{Hild, LHD}, Nitsche~\cite{Nitsche}, and Chen \& Fried~\cite{CF}, the spanning surface is described by a smooth, bijective mapping of the unit disk onto a surface embedded in three-dimensional space, the boundary of which coincides with centerline of the loop. In Section~\ref{equilibrium}, the linearized versions of the general equilibrium conditions are derived 
and an analysis of bifurcations from a flat, circular ground state is performed for different ranges of the salient dimensionless parameters. {Physical interpretations of the results are provided} in Section~\ref{discussion} and concluding remarks appear in Section~\ref{conclusion}.

\section{Preliminaries}
\label{prelim}

Consider a closed flexible filament
spanned by a soap film. Let the filament be modeled as an inextensible, unshearable, isotropic, and linearly-elastic rod with uniform circular cross-section and centerline $\calC$. Take the flexural rigidity $a>0$ and torsional rigidity $c>0$ of the rod to be uniform. Let the soap film be modeled as a  surface $\calS$ with uniform tension $\sigma>0$. Assume that $\calC$ is free of self-contact, that $\calS$ is orientable, and that the boundary of $\calS$ and $\calC$ coincide:
\be
\partial\calS=\calC.
\label{coincidence}
\ee

In the context of special Cosserat theory of rods, as presented by Antman~\cite{antman}, the configuration of a rod is characterized by its centerline $\calC$ and a triad $\{\bfd_1,\bfd_2,\bfd_3\}$ of orthonormal directors. Whereas $\bfd_1$ and $\bfd_2$ reside in the normal cross-section of the rod, $\bfd_3$ is perpendicular to that cross section. For an inextensible, unshearable, isotropic rod, the third director $\bfd_3$ and the unit tangent vector $\bft$ of the Frenet frame of $\calC$ must be equal, giving 
\be
\bfd_3=\bft.
\label{directors0}
\ee
Additionally, the remaining directors $\bfd_1$ and $\bfd_2$ must lie in the plane spanned by the normal $\bfn$ and binormal $\bfb$ of the Frenet frame of $\calC$ and, thus, be related to $\bfn$ and $\bfb$ by rotation through some angle $\psi$, so that
\be
\left.
\begin{split}
\bfd_1&=\phantom{-}(\cos\psi)\bfn+(\sin\psi)\bfb,
\\[4pt]
\bfd_2&=-(\sin\psi)\bfn+(\cos\psi)\bfb.
\end{split}
\,\right\}
\label{directors}
\ee

The curvature $\kappa$, torsion $\tau$, and twist density $\Omega$ of $\calC$ are given by
\be
\kappa=|\bft'|,
\qquad
\kappa^2\tau=(\bft'\times\bft^{\prime\prime})\cdot\bft^{\prime\prime\prime},
\qquad\text{and}\qquad
\Omega=(\bfd_1\times\bfd_1^\prime)\cdot\bfd_3,
\label{Omegakappatau}
\ee
where a prime is used to denote differentiation with respect to arclength $s$ along $\calC$.  By \eqref{directors} and the Frenet--Serret relations
\be
\bft'=\kappa\bfn,
\qquad
\bfn'=-\kappa\bft+\tau\bfb, 
\qquad
\bfb'=-\tau\bfn,
\label{SF}
\ee
the arclength derivative of $\bfd_1$ can be expressed as $\bfd'_1=(\tau+\psi')\bfd_2-(\kappa\cos\psi)\bft$,
which, with \eqref{Omegakappatau}${_3}$,  yields
\be
\Omega=\tau+\psi'.
\label{twist}
\ee
As a consequence of \eqref{twist}, the twist density $\Omega$ may be nontrivial even if the torsion $\tau$ vanishes; moreover, $\Omega$ generally differs from $\tau$ unless the angle $\psi$ needed to rotate the normal $\bfn$ and binormal $\bfb$ of the Frenet frame into the cross-sectional directors $\bfd_1$ and $\bfd_2$ does not vary along the centerline $\calC$.

\section{Net potential-energy}
\label{generalformulation}
In view of the foregoing discussion and granted that gravitational effects are negligible, the net potential energy $F$ of the system comprised by the flexible loop and the soap film takes the form
\be
F=\int_{\calC}\half(a\kappa^2+\crig\Omega^2)+\int_{\calS}\sigma.
\label{EL1}
\ee
In the degenerate case $c=0$ of vanishing twisting rigidity, \eqref{EL1} reduces to the net potential-energy of the Euler--Plateau problem (Giomi \& Mahadevan~\cite{gm}; Chen \& Fried~\cite{CF}).

Consider a flat, circular configuration of the system with radius $R>0$ and uniform twist density $\Omega$. For such a configuration, \eqref{EL1} specializes to
\be
\bigg(1+\frac{a}{c}\bigg(\frac{R\Omega c}{a}\bigg)^{\!\!2}+\frac{R^3\sigma}{a}\bigg)\frac{\pi a}{R}.
\label{Fground}
\ee
Aside from demonstrating that the net free-energy of a flat, circular ground state of radius $R$ scales with the lineal bending-energy $\pi a/R$ of the boundary, \eqref{Fground} reveals the potential significance of three dimensionless parameters:
\be
\chi=\frac{c}{a}>0,
\qquad
\mu=\frac{R\Omega c}{a},
\qquad
\nu=\frac{R^3\sigma}{a}>0.
\label{dgroups1}
\ee
Whereas $\chi$ simply measures the importance of the twisting rigidity $c$ relative to the flexural rigidity $a$, $\mu$ is an amalgamation {between} the competing influences encoded in $\chi$ and the dimensionless twist density $R\Omega$. Finally, $\nu$, which is familiar from the Euler--Plateau problem, accounts for the importance of the areal free-energy $\pi R^2\sigma$ relative to the lineal bending-energy $\pi a/R$ (Chen \& Fried \cite{CF}).

\section{Parametrization}
\label{parameter}
Suppose that $\calS$ is parametrized in accord with
\be
\calS=\{\bfx\in\mathbb{R}^3:\bfx=R\mskip2mu\bfxi(r,\theta), 0\le r\le1, 0\le \theta \le 2\pi\},
\label{Srep}
\ee
where $R$ is the radius of a circle with perimeter equal to the length of $\calC$, $\bfxi$ is a four-times continuously differentiable, injective mapping defined on the closed unit disk, the dimensionless radius $r$ and the polar angle $\theta$ provide polar coordinates on the closed unit disk with associated basis vectors $\er$ and $\et$. The smoothness of $\bfxi$ implies that $\bfxi({r},\cdot)$ must be periodic on $[0,2\pi]$ for each $r$ in $(0,1]$, as must its derivatives with respect to $\theta$ up to the fourth order; moreover, $\bfxi$ must satisfy closure conditions
\be
\bfxi({r},0)=\bfxi(r,2\pi),
\qquad
\xit({r},0)=\xit(r,2\pi),
\qquad
0<{r}\le1,
\ee
ensuring that it smoothly maps each concentric circle belonging to the unit disk into a closed curve on $\calS$ and, moreover, that two distinct values of $r$ are mapped to curves on $\calS$ that do not intersect.

With reference to \eqref{Srep}, the assumption \eqref{coincidence} that $\partial\calS$ and $\calC$ coincide induces a parametrization
\be
\calC=\{\bfx\in\mathbb{R}^3:\bfx=R\bfxi(1,\theta),\,0\le\theta\le2\pi\}
\label{Crep}
\ee
of $\calC$, where, to ensure satisfaction of the inextensibility constraint, $\bfxi$ must be consistent with
\be
|\xit|_{\reqone}=1.
\label{Pr1}
\ee
As consequences of \eqref{Crep} and \eqref{Pr1}, the curvature, torsion, and twist density of $\calC$ may be expressed as
\be
\kappa=\frac{|\xitt|}{R}\bigg|_{\reqone},
\quad 
\tau=\frac{(\xit\times\xitt)\cdot\xittt}{R|\xitt|^2}\bigg|_{\reqone}, 
\quad\text{and}\quad
\Omega=\frac{(\xit\times\xitt)\cdot\xittt}{R|\xitt|^2}\bigg|_{\reqone}
+\frac{\psi_\theta}{R},
\label{kappatauomegaparrep}
\ee
from which it follows that the lineal contribution to the net potential-energy $F$ defined in \eqref{EL1} admits a parametrized representation of the form
\be
\int_{\calC}\half(a\kappa^2+\crig\Omega^2)=\frac{a}{R}
\lint\frac{1}{2}\bigg[|\xitt|^2+\chi\bigg(\frac{(\xit\times\xitt)\cdot\xittt}{|\xitt|^2}
+\psi_\theta\bigg)^{\!\!2}\,\bigg]_{\reqone}\!\!\dtheta,
\label{Fl}
\ee
with $\chi>0$ as defined in \eqref{dgroups1}$_1$. Further, since the area of a surface element of $\calS$ spanned by $\text{d}r$ and $\text{d}\theta$ is simply $|\xir\dr\ctimes\xit\dtheta|=|\xir\ctimes\xit|\drdtheta$, the areal contribution to $F$ can be expressed parametrically as
\be
\int_{\calS}\sigma=\frac{a}{R}\aintdim\nu|\xir\times\xit|\drdtheta,
\label{Fa}
\ee
with $\nu>0$ as defined in \eqref{dgroups1}$_3$.

Like the net potential-energy \eqref{Fground} of a flat, circular configuration with given uniform twist density, the lineal and areal free energies \eqref{Fl} and \eqref{Fa} scale with $a/R$. It is therefore convenient to work with the dimensionless net potential-energy $\Phi$ defined by
\be
\Phi=\frac{RF}{a}
=\lint\frac{1}{2}\bigg[|\xitt|^2+\chi\bigg(\frac{(\xit\times\xitt)\cdot\xittt}{|\xitt|^2}
+\psi_\theta\bigg)^{\!\!2}\,\bigg]_{\reqone}\!\!\dtheta
+\aintdim\nu|\xir \times \xit| \drdtheta.
\label{Pr2}
\ee

\section{Equilibrium conditions}
\label{equilibrium}

At equilibrium, the first-variation condition
\be
\del\Phi=0
\label{fvc}
\ee
must necessarily hold. Consider smooth variations $\bfw=\del\bfxi$ and $\iota=\del\psi$ of $\bfxi$ and $\psi$. Prior to computing $\del\Phi$, consider the implications of the constraint \eqref{Pr1}. In particular, on differentiating \eqref{Pr1}, it follows that $\xit$ and $\wt$ must jointly satisfy
\be
[\xit\cdot\wt]_{\reqone}=0.
\label{Pr3}
\ee
Further, $(\del\psi)_\theta=\del(\psi_\theta)$, from which it follows that
\be
\iota_\theta=\del(\psi_\theta).
\label{Pr6}
\ee
An important consequence of \eqref{Pr3} is that, for any vector field $\bfp$ defined on $r=1$ and consistent with the condition $\bfp(0)=\bfp(2\pi)$,
\be
[\bfp\cdot\wt]_{\reqone}=[(\bfp-q\xit)\cdot\wt]_{\reqone},
\label{Pr3p}
\ee
where $q$ is a scalar field defined on $r=1$ and consistent with $q(0)=q(2\pi)$.

Following Chen \& Fried \cite{CF}, varying the sum of the flexural and areal contributions to $\Phi$ yields
\begin{multline}
\del\bigg(\lint\half|\xitt|^2\big|_{\reqone}\dtheta
+\aintdim\nu|\xir\times\xit|\drdtheta\bigg)
\\[4pt]
=\lint[(\nu\xit\times\bfm+(\xittt-\lambda_1\xit))_{\theta}\cdot\bfw]_{\reqone}\dtheta
\\
-\aintdim \nu(\xit\times\bfm_{ r}+\bfm_\theta\times\xir)\cdot\bfw\drdtheta,
\label{Pr4+5}
\end{multline}
where $\bfm$ defined according to
\be
\bfm=\frac{\xir\times\xit}{|\xir\times\xit|}
\label{mparrep}
\ee
provides a unit orientation for $\calS$ and, bearing in mind \eqref{Pr3p}, $\lambda_1$ is a Lagrange multiplier needed to ensure satisfaction of the constraint \eqref{Pr1} of inextensibility.

Next, consider the twist contribution to \eqref{Pr2}. By \eqref{kappatauomegaparrep}$_2$ and \eqref{Pr3p}, 
\begin{align}
\del\bigg(
\frac{(\xit\times\xitt)\cdot\xittt}{|\xitt|^2}\bigg|_{\reqone}
\bigg)
&=\bigg[\frac{ (\wt \times \xitt) \cdot \xittt +(\xit \times \wtt) \cdot \xittt +(\xit \times \xitt) \cdot \wttt}{|\xitt|^2} 
\notag\\[4pt]
&\qquad\qquad\qquad\qquad\qquad-\frac{2((\xit \times \xitt) \cdot \xittt) (\xitt \cdot \wtt)}{|\xitt|^4}\bigg]_{\reqone}
\notag\\[8pt]
&= \bigg[\bigg( \frac{ \xitt \times \xittt}{|\xitt|^2}-\lambda_2\xit \bigg) \cdot \wt
+\bigg(\frac {(\xit \times \xitt) \cdot \wtt}{|\xitt|^2}\bigg)_{\!\!\theta}\mskip2mu\bigg]_{\reqone},
\label{Pr7}
\end{align}
where $\lambda_2$ is another Lagrange multiplier required by \eqref{Pr1}. In view of \eqref{Pr3p} and \eqref{Pr7}, varying the twist contribution to \eqref{Pr2} yields
\begin{multline}
\del\lint\frac{\chi}{2}\bigg(\frac{(\xit\times\xitt)\cdot\xittt}{|\xitt|^2}
+\psi_\theta\bigg)^{\!\!2}\bigg|_{\reqone}\dtheta
\\[8pt]
=\lint\chi\bigg[\bigg(\frac{(\xit\times\xitt)\cdot\xittt}{|\xitt|^2}
+\psi_\theta\bigg)\bigg( \frac{ \xitt \times \xittt}{|\xitt|^2}-\lambda_2\xit \bigg) \cdot\wt
\bigg]_{\reqone}\dtheta
\\[4pt]
-\lint\chi\bigg[
\bigg(\frac{(\xit\times\xitt)\cdot\xittt}{|\xitt|^2}
+\psi_\theta\bigg)_{\!\!\theta}\frac{(\xit\times\xitt)\cdot\wtt}{|\xitt|^2}\bigg]_{\reqone}\dtheta
\\[4pt]
-\lint\chi\bigg[
\bigg(\frac{(\xit\times\xitt)\cdot\xittt}{|\xitt|^2}
+\psi_\theta
\bigg)_{\!\!\theta}\mskip2mu\bigg]_{\reqone}\iota\dtheta.
\label{Pr8}
\end{multline}

Further, combining \eqref{Pr4+5} and \eqref{Pr8} yields 
\begin{multline}
\del\Phi=\lint\bigg\{\nu\xit\times\bfm
+\bigg[\bigg(\xittt
-\chi\bigg(\frac{(\xit\times\xitt)\cdot\xittt}{|\xitt|^2}
+\psi_\theta\bigg)\frac{\xitt\times\xittt}{|\xitt|^2}
-\varsigma\xit\bigg)_{\!\!\theta}\bigg]\cdot\bfw\bigg\}_{\reqone}\dtheta
\\[8pt]
+\lint\chi
\bigg[\bigg(\bigg(\frac{(\xit\times\xitt)\cdot\xittt}{|\xitt|^2}
+\psi_\theta
\bigg)_{\!\!\theta}\frac{\xit\times\xitt}{|\xitt|^2}\bigg)_{\!\!\theta}\cdot\wt \bigg]_{\reqone}\dtheta
\\[8pt]
-\lint\chi\bigg[
\bigg(\frac{(\xit\times\xitt)\cdot\xittt}{|\xitt|^2}+\psi_\theta
\bigg)_{\!\!\theta}\mskip2mu\bigg]_{\reqone}\iota\dtheta
\\[8pt]
-\aintdim\nu(\xit\times\bfm_r+\bfm_\theta\times\xir)\cdot\bfw\drhodtheta,
\label{Pr9}
\end{multline}
where $\lambda$ is a composite Lagrange multiplier formed by a difference,
\be
\lambda=\lambda_1-\frac{c}{a}\bigg(\frac{(\xit\times\xitt)\cdot\xittt}{|\xitt|^2}
+\psi_\theta\bigg)\lambda_2
=\lambda_1-\mu\lambda_2,
\label{cmultiplier}
\ee
involving the  dimensionless group $\mu$ defined in \eqref{dgroups1}$_2$ and Lagrange multipliers $\lambda_1$ and $\lambda_2$ entering \eqref{Pr4+5} and \eqref{Pr8}. 

For $\bfw$ compactly supported on the interior of the unit disk, applying the first-variation condition \eqref{fvc} to \eqref{Pr9} yields the areal equilibrium condition 
\be
(\xit\times\bfm_r+\bfm_\theta\times\xir)\cdot\bfm=0,
\label{Pr12}
\ee
which imposes force balance in the direction normal $\bfm$ to $\calS$.

Further, since $\bfw$ and $\iota$ may be chosen independently on the boundary of the unit disk, \eqref{fvc} and \eqref{Pr9} yield two lineal equilibrium conditions. Whereas one of these conditions,
\be
R\Omega=\frac{(\xit\times\xitt)\cdot\xittt}{|\xitt|^2}\bigg|_{\reqone}+
\psi_\theta
=\text{constant},
\label{Omegat}
\ee
dictates that the twist density of $\calC$ must be uniform, the remaining condition,
\be
\bigg[\nu\go\xit\times\bfm+\bigg(\xittt-\mu\frac{\xitt\times\xittt}{|\xitt|^2}-\lambda\go\xit\bigg)_{\!\!\theta}\mskip2mu\bigg]_{\reqone}={\bf0},
\label{nd4}
\ee
expresses force balance on $\calC$.

Since the areal contribution to the net free-energy considered here is identical to that of the Euler--Plateau problem (without twist), it is not surprising that \eqref{Pr12} is equivalent to the dimensionless areal equilibrium condition of Chen \& Fried \cite[eq.~(100)]{CF}, who denote the unit normal to $\calS$ by $\bfn$ instead of $\bfm$. Additionally, {in the degenerate case where} the twisting rigidity $c$ obeys $c=0$ (so that, by \eqref{dgroups1}$_2$, $\mu=0$), \eqref{nd4} reduces to the lineal equilibrium condition of the Euler--Plateau problem obtained by Chen \& Fried \cite{CF}. 

It is noteworthy that Chen \& Fried \cite{CF} establish the equivalence between the equilibrium conditions obtained by Giomi \& Mahadevan \cite{gm} and those arising in the parametrized setting. Chen \& Fried~\cite{CF} also show that the areal equilibrium condition is equivalent to the requirement,
\be
H=\frac{(\abs{\xit}^2\xirr-2(\xir \cdot \xit)\xirt+\abs{\xir}^2\xitt)\cdot\bfm}{2R \abs{\xir\times\xit}^2}
=0,
\label{Hparrep}
\ee
that the mean curvature of $\calS$ vanishes pointwise. 

To acquire some insight regarding the geometric content of lineal equilibrium condition \eqref{nd4}, note that arclength on $r=1$ is measured by
\be
s=R\theta, 
\qquad
0\le\theta\le2\pi,
\label{arclength}
\ee
and define $\bfz$ such that
\be
\bfz(s)=R\bfxi(1,R^{-1}s),
\qquad
0\le s\le2\pi R.
\label{barx}
\ee
Then, by the Frenet--Serret relations \eqref{SF} and the parametric representations \eqref{kappatauomegaparrep} of $\kappa$, $\tau$, and $\Omega$, the lineal equilibrium condition \eqref{nd4} can be expressed as
\be
a(\kappa\bfn)''-(c\Omega\bfn\times\bfn'+\lambda\bft)'+\sigma\bfnu=\bf0,
\ee
or, equivalently, {as a system,}
\be
\left.
\begin{array}{c}
\displaystyle
\lambda+\frac{3}{2} a \kappa^2+\crig\Omega\tau= \beta,
\cr\noalign{\vskip8pt}\displaystyle
\kappa''+\half \kappa^3-\kappa\bigg(\tau^2-\frac{\crig\Omega}{a}\tau+\frac{\beta}{a}\bigg)-\frac{\sigma}{a}\sin{\vartheta}=0,
\cr\noalign{\vskip8pt}\displaystyle
\kappa'\bigg(2\tau-\frac{\crig \Omega}{a}\bigg)+\kappa\tau'+\frac{\sigma}{a} \cos{\vartheta}=0,
\end{array}
\right\}
\label{ELpr}
\ee
{of three scalar equations,} with $\beta$ a constant. In the degenerate case $c=0$ of vanishing twisting rigidity, \eqref{ELpr}$_{2,3}$ reduce to the lineal equilibrium conditions of Giomi \& Mahadevan~\cite[eqs.~(2.8$a$, $b$)]{gm}. 

\section{Solution to the linearized equilibrium conditions}
\label{linear}

As observed by Chen \& Fried \cite{CF}, a planar disk bounded by a circle of unit radius provides a trivial solution to the dimensionless version of the Euler--Plateau problem. For this  solution, $\bfxi$ and $\lambda$ are given by
\be
\bfxi({r},\theta)=r\er(\theta)
\qquad\text{and}\qquad
\lambda=-(1+\nu).
\label{ts1}
\ee

Since $\bfe_\theta=\et$ and $\bfe_{\theta\theta}=-\bfe$, it follows that
\be
\xir=\er, \qquad \xit={r}\et, \qquad \xitt=-{r}\er, \qquad \xittt=-{ r}\et,  \qquad \xitttt={r}\er,
\label{ts2}
\ee
Further, by {\eqref{ts2}$_{\text{2--4}}$,}
\be
(\xit\times\xitt)\cdot\xittt=0,
\ee
whereby the twist contribution to the dimensionless lineal equilibrium condition \eqref{nd4} vanishes. Thus, \eqref{ts1} also provides a trivial solution to the boundary-value problem \eqref{Pr12}--\eqref{nd4}. 

Consider perturbations
\be
\bfxi(r,\theta)=r\bfe+\bfeta
\qquad \text{and} \qquad
\lambda(\theta)=-(1+\nu)+\epsilon(\theta)
\label{perturbations}
\ee
of \eqref{ts1} with $\bfu$ and $\epsilon$ sufficiently small. Straightforward but lengthy calculations then lead to linearized versions,
\be
\bigg(\urro+\frac{1}{r} \uro+\frac{1}{r^2}\utt\bigg)\cdot(\ez)=0
\label{lin1}
\ee
and
\begin{multline}
[\utttt+(1+\nu)\utt+\big(\nu(\uro \cdot \ez)-2\mu(\utt \cdot \er)_\theta \big)\ez
\\[4pt]
+\nu \ut \times (\ez)
+\mu\er \times(\utttt+\utt)+\epsilon \er-\epsilon_\theta \et]_{\reqone}=\bf0,
\label{lin2}
\end{multline}
of the equilibrium equations \eqref{Pr12} and \eqref{nd4}. Although it is also possible to consider a perturbation of $\mu$, doing so does not alter the results up to the order of significance considered here.

The perturbation $\bfeta$ entering \eqref{perturbations}$_1$ admits a decomposition of the form 
\be
\bfeta=v\er+u\et+w\ez,
\label{decomp}
\ee
subject to the linearized version
\be
[\ut\cdot\et]_{\reqone}=[\uit+v]_{\reqone}=0
\label{ts3}
\ee
of the constraint \eqref{Pr3}. In view of \eqref{decomp} and \eqref{ts3}, the restriction of $\ut$ to ${r}=1$ must obey 
\be
\ut|_{\reqone}=[\vit-u]_{\reqone}\er+\wit|_{\reqone}\mskip2mu\ez,
\label{ts4}
\ee
repeated  differentiation with respect to $\theta$ of which yields {the following} useful identities:
\be
\left.
\begin{split}
\utt|_{\reqone}&=[\vitt+v]_{\reqone}\mskip2mu\er+[\vit-u]_{\reqone}\mskip2mu\et+\witt|_{\reqone}\mskip2mu\ez, 
\\[4pt]
\uttt|_{\reqone}&=[\vittt+u]_{\reqone}\mskip2mu\er+2[\vitt+v]_{\reqone}\mskip2mu\et+\wittt|_{\reqone}\mskip2mu\ez,
\\[4pt]
\utttt|_{\reqone}&=[\vitttt-2\vitt-3v]_{\reqone}\mskip2mu\er+[3\vittt+2\vit+u]_{\reqone}\mskip2mu\et
+\witttt|_{\reqone}\mskip2mu\ez.
\end{split}
\right\}
\label{ts5to7}
\ee

Substituting \eqref{decomp} in the linear equilibrium equation \eqref{lin1} leads to the scalar Laplace equation,
\be
\wirr+\frac{1}{r}\wir+\frac{1}{r^2}\witt=0,
\label{lin5}
\ee
for the {transverse} perturbation $w$. Similarly, substituting the decomposition \eqref{decomp}, the inextensibility constraint \eqref{ts3}, and the results \eqref{ts4} and \eqref{ts5to7} in the linearized boundary condition \eqref{lin2}, and subsequently decomposing it into its  transverse, radial, and azimuthal components results in three scalar linear equilibrium equations
\be
\left.
\begin{array}{c}
[  \witttt+(1+\nu)\witt+\nu \wir +\mu (\vitt+v)_{\theta} ]_{\reqone}=0,
\cr\noalign{\vskip6pt}\displaystyle
[ \vitttt-(1-\nu)(\vitt+v)-v+\epsilon]_{\reqone}=0,
\cr\noalign{\vskip6pt}\displaystyle
[ 3(\vitt+v)_{\theta}  - \mu (\witt+w)_{\theta \theta}-\epsilon_\theta]_{\reqone}=0.
\end{array}
\right\}
\label{lin6}
\ee 

Assuming that \eqref{lin5} admits separable solutions of the form 
\be
w(r,\theta)=\varrho(r)\Theta(\theta),
 \label{lin9}
\ee
with $\varrho$ bounded and $\Theta$ periodic, leads to solutions of the form
\be
w(r,\theta)=r^n (a_n \cos n\theta + b_n \sin n\theta), 
\qquad n\in\{0,1,2,\dots\}.
\label{lin11}
\ee
Next, integrating \eqref{lin6}$_3$ determines the perturbation $\epsilon$ entering \eqref{perturbations}$_2$ through 
\be
\epsilon=[3(\vitt+v)-\mu(\wittt+\wit)]_{\reqone}+C,
\label{lin12}
\ee
with $C$ a constant. Using \eqref{lin12} in \eqref{lin6}$_2$ gives 
\be
[\vitttt+(\nu+2)(\vitt+v) -v - \mu(\wittt+\wit)]_{\reqone}+C=0.
\label{lin13}
\ee
Being periodic, the restriction $v |_{\reqone}$ of the radial perturbation $v$ to $r=1$ admits a representation of the form  
\be
v(1,\theta)=c_n \cos n\theta + d_n \sin n\theta,  
\qquad n\in\{0,1,2,\dots\},
\label{lin14}
\ee
where in view of \eqref{lin13}, $c_0=C/(1+\nu)$. As a consequence of \eqref{lin14}, the inextensiblity constraint \eqref{ts3} requires that the azimutional perturbation $u$ obeys
\be
u(1,\theta)=\frac{c_n \sin n\theta - d_n \cos n\theta}{n},  \qquad n\in\{1,2,3,\dots\}.
\label{lin15}
\ee

\begin{figure}
\includegraphics[width=4.4 in]{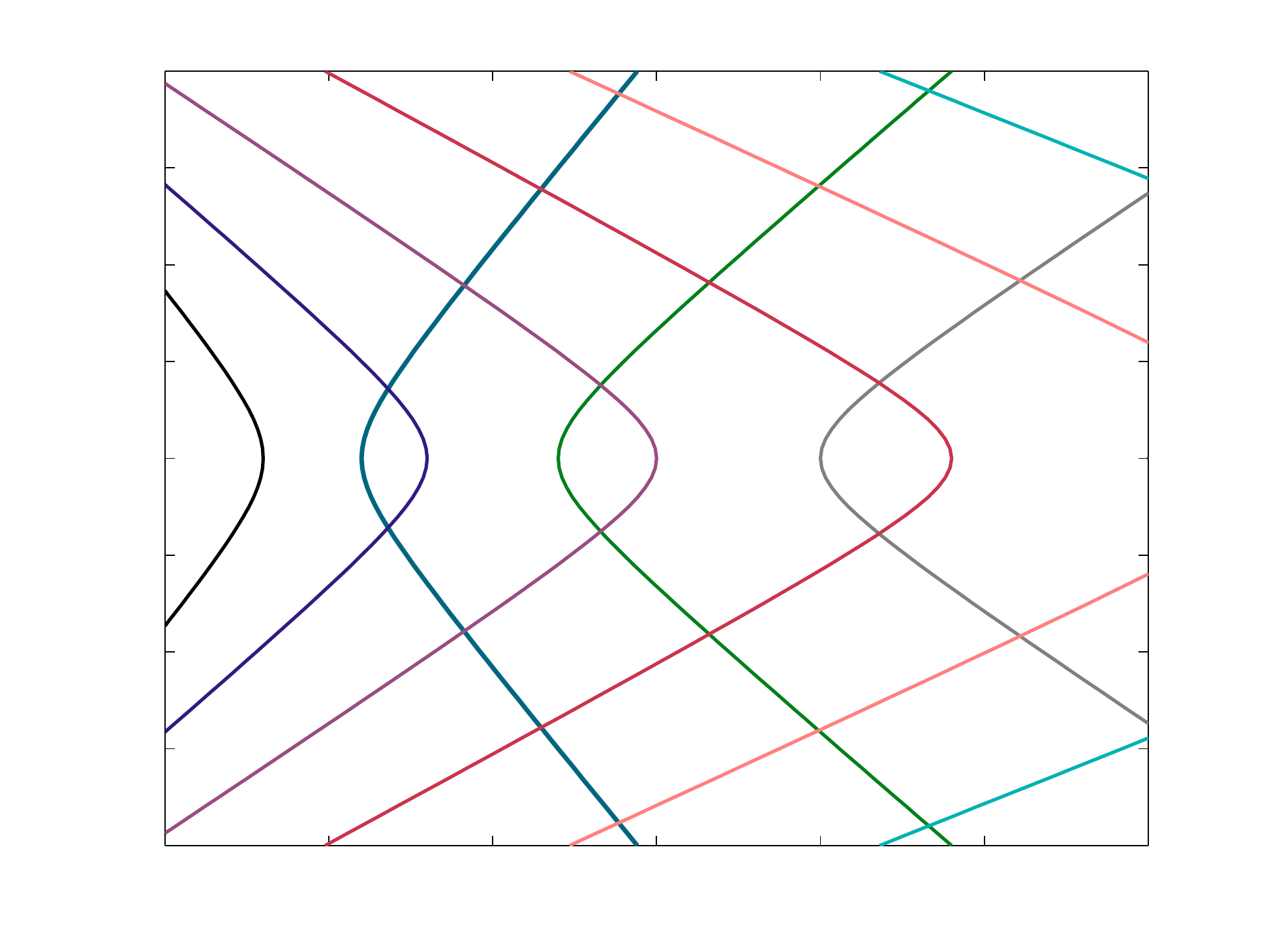}
\put(-305,120){\large$\mu$}
\put(-158,0){\large$\nu$}
\put(-285,216.5){$4$}
\put(-285,192.5){$3$}
\put(-285,168.5){$2$}
\put(-285,144.5){$1$}
\put(-285,120.5){$0$}
\put(-292,94){$-1$}
\put(-292, 70){$-2$}
\put(-292, 46){$-3$}
\put(-292, 22){$-4$}
\put(-277,15){$0$}
\put(-237,15){$5$}
\put(-199,15){$10$}
\put(-158,15){$15$}
\put(-117,15){$20$}
\put(-76,15){$25$}
\put(-35,15){$30$}
\put(-20,216.5){$n=2$ negative-branch \textcolor[rgb]{0,0,0}{\large---}}
\put(-20,192.5){$n=2$ positive-branch \textcolor[rgb]{0,0.4,0.5}{\large---}}
\put(-20,168.5){$n=3$ negative-branch \textcolor[rgb]{0.2,0.2,0.5}{\large---}}
\put(-20,144.5){$n=3$ positive-branch \textcolor[rgb]{0,0.5,0.1}{\large---}}
\put(-20,120.5){$n=4$ negative-branch \textcolor[rgb]{0.6,0.3,0.5}{\large---}}
\put(-20,94){$n=4$ positive-branch \textcolor[rgb]{0.5,0.5,0.5}{\large---}}
\put(-20, 70){$n=5$ negative-branch \textcolor[rgb]{0.8,0.2,0.3}{\large---}}
\put(-20, 46){$n=6$ negative-branch \textcolor[rgb]{1,0.5,0.5}{\large---}}
\put(-20, 22){$n=7$ negative-branch \textcolor[rgb]{0,0.7,0.7}{\large---}}
\vspace{-4pt}
\caption{
Two-parameter bifurcation diagram for the system of linear equilibrium equations. While $\mu$ measures the importance of the twisting moment of the bounding loop relative to the bending moment of the bounding loop, $\nu$ indicates the significance of surface tension relative to lineal bending.}
\label{f1}
\end{figure}

Additionally, for each $n\ge1$, substituting \eqref{lin11} and \eqref{lin14} in the linearized boundary conditions \eqref{lin6}$_1$ and \eqref{lin13} yields a pair,
\be
\left[ 
\begin{array}{cc}
-n(n^2-1)\mu & n(n-1)(n(n+1)-\nu)
\cr\noalign{\vskip6pt}
(n^2-1)(n^2-(\nu+1)) & -n(n^2-1)\mu
\end{array}
\right]
\left[ 
\begin{array}{cc}
a_n
\cr\noalign{\vskip6pt} 
d_n
\end{array}  
\right]=
\left[
\begin{array}{cc}
0
\cr\noalign{\vskip6pt} 
0
\end{array}
\right]
 \label{lin16}
 \ee
and 
\be
\left[
\begin{array}{cc}
n(n^2-1)\mu & n(n-1)(n(n+1)-\nu)
\cr\noalign{\vskip6pt} 
(n^2-1)(n^2-(\nu+1)) & n(n^2-1)\mu
\end{array}
\right]
\left[
\begin{array}{cc}
b_n
\cr\noalign{\vskip6pt} 
c_n
\end{array}
\right]=
\left[
\begin{array}{cc}
0
\cr\noalign{\vskip6pt} 
0
\end{array}
\right],
\label{lin15}
\ee
of homogeneous linear systems involving the coefficients $a_n$, $b_n$, $c_n$, and $d_n$. For \eqref{lin16} and \eqref{lin15} to possess nontrivial solutions, the determinants of the relevant coefficient matrices must vanish. Since those matrices differ only by the signs of their diagonal elements, their determinants are equal. A quick calculation shows that the solvability condition takes the form
\be
n(n-1)(n^2-1)[\nu^2-(n+1)(2n-1)\nu-n(n+1)(\mu^2-n^2+1)]=0.
\label{lin17}
\ee
The choice $n=1$ describes rigid body rotations and, hence, is of no physical interest. For $n\ge2$, \eqref{lin17} holds if and only if $\nu$ and $\mu$ satisfy
\be
\nu^2-(n+1)(2n-1)\nu+n(n+1)(n^2-1-\mu^2)=0.
\label{lin18}
\ee
If $n$ and $\mu$ are viewed as given, \eqref{lin18} is a quadratic equation for $\nu$. By \eqref{dgroups1}$_3$, $\nu$ must be positive and is thus determined by
\be
\nu=\left\{
\begin{array}{l l}
\displaystyle
 (n+1)\bigg(n-\frac{1}{2}\pm \frac{1}{2}\sqrt{1+\frac{4n\mu^2}{n+1}}\,\bigg),   &\quad  \text{if} \quad \mu^2+1<n^2,
\cr\noalign{\vskip6pt}\displaystyle
(n+1)\bigg(n-\frac{1}{2} + \frac{1}{2}\sqrt{1+\frac{4n\mu^2}{n+1}}\,\bigg), & \quad \text{otherwise}.
\end{array}
\right.
 \label{lin19}
 \ee
These bifurcation points are depicted in Figure~\ref{f1}.

\begin{figure}
\centering
\includegraphics[width=4 in]{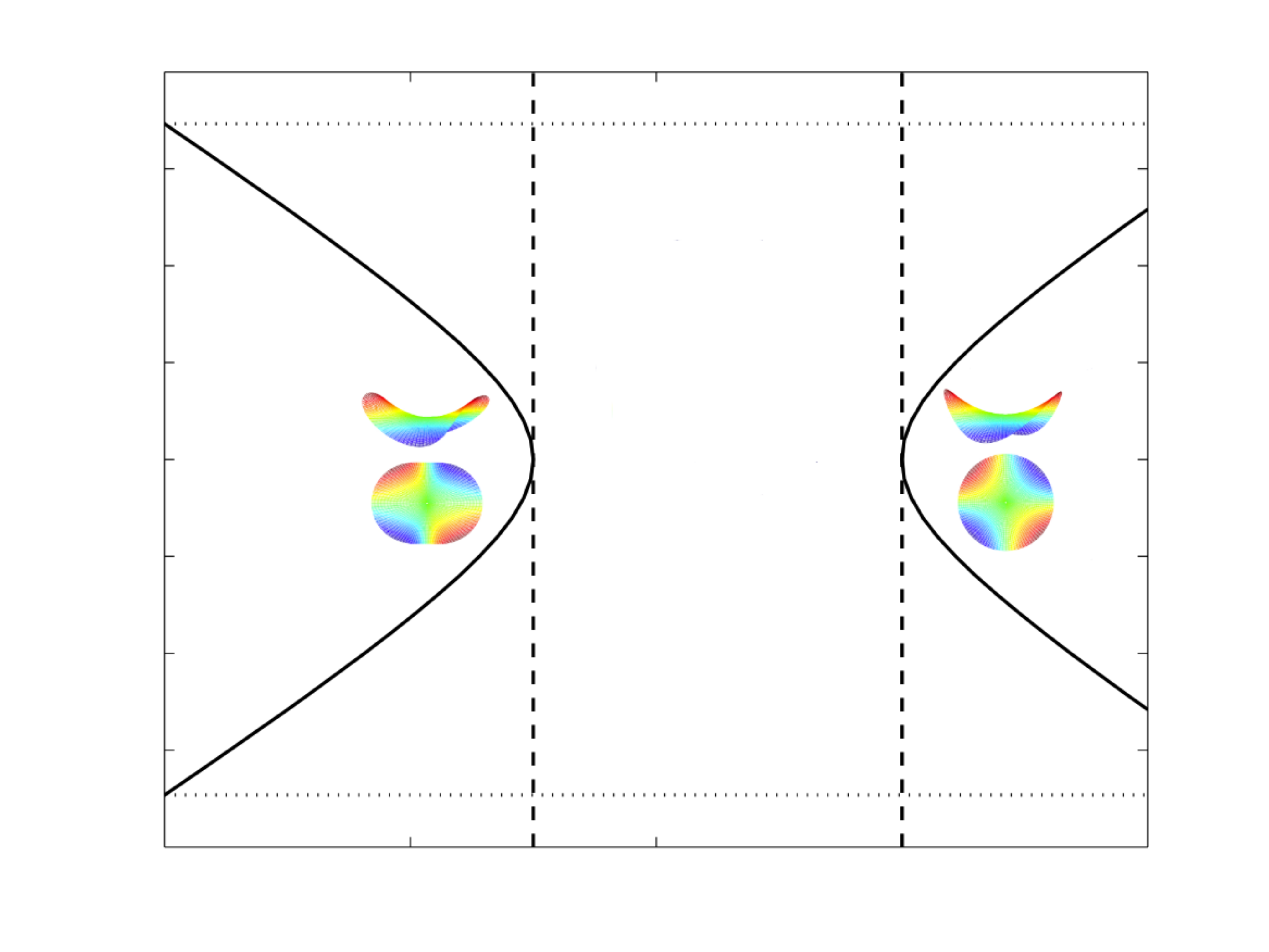}
\put(-286,110){\large$\mu$}
\put(-142,-6){\large$\nu$}
\put(-260,197){$2$}
\put(-260,153){$1$}
\put(-260,109){$0$}
\put(-267,64){$-1$}
\put(-267, 20){$-2$}
\put(-254,13){$0$}
\put(-198,13){$2$}
\put(-142,13){$4$}
\put(-86,13){$6$}
\put(-30,13){$8$}
\caption{Comparison of the bifurcation parameters in the Euler--Plateau problem with twist~(---), the classical Euler--Plateau problem (- - -), and the Michell (.....) problem. For $\mu=0$, the bifurcation criterion specializes to that of the Euler--Plateau problem (Giomi \& Mahadevan~\cite{gm}, Chen \& Fried~\cite{CF}), whereas for $\nu=0$ the critical twist of Michell~\cite{Michell} is recovered. Resistance to twist has an stabilizing effect on out-of-plane (saddle) bifurcations from a circular ground state in the Euler--Plateau problem. Nevertheless, a twisted, elastic, circular ring becomes less stable when spanned by a soap film. 
}
\label{f3}
\end{figure}
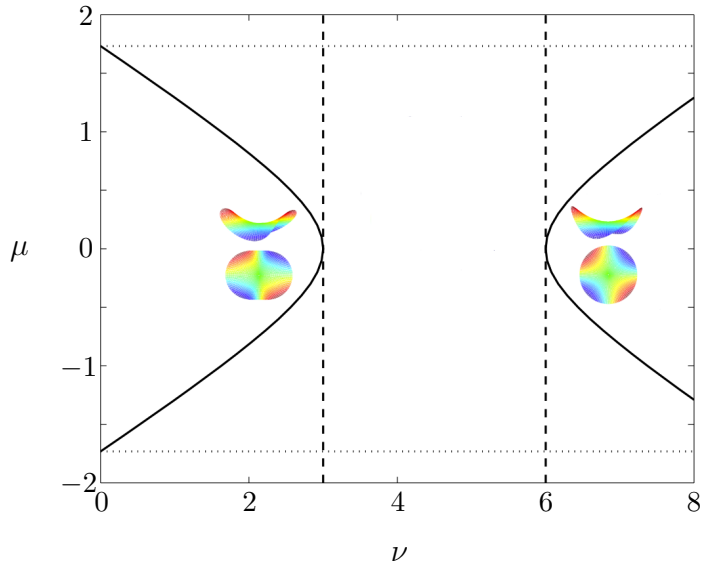
%

On neglecting twist and, thus, setting $\mu=0$, \eqref{lin19} simplifies to 
\be
\nu=(n+1)(n-\half \pm \half),
 \label{lin20}
 \ee
which gives $\nu=n^2-1$ and $\nu=n(n+1)$, consistent with the results of Chen \& Fried \cite{CF} for {in-plane} and {transverse} bifurcations. In the absence of twist, the first bifurcations are therefore triggered at $\nu=3$ and $\nu=6$, corresponding to $n=2$, as depicted in Figure~\ref{f3}.

Another interesting specialization arises on {neglecting the surface tension $\sigma$ of the soap film, so that, by \eqref{dgroups1}$_3$, $\nu=0$ and} \eqref{lin17} reduces to 
\be
\mu^2=n^2-1.
\label{lin21}
\ee
In combination with the definition \eqref{dgroups1}$_2$ of $\mu$, \eqref{lin21} yields the well-known critical twist density $\Omega=(n^2-1)a/Rc$ for the bifurcation of twisted annular rings, which was obtained independently by Michell~\cite{Michell}, Zajac~\cite{Zajac}, and Benham~\cite{Benham}. In particular, \eqref{lin21} demonstrates that a twisted ring first buckles at $\mu=\sqrt{3}$, corresponding to the mode $n=2$. This phenomenon has been termed the \emph{Michell instability} by Goriely~\cite{Goriely}, who also describes the salient historical developments.  
\begin{figure}
\centering
\includegraphics [width=137 mm]{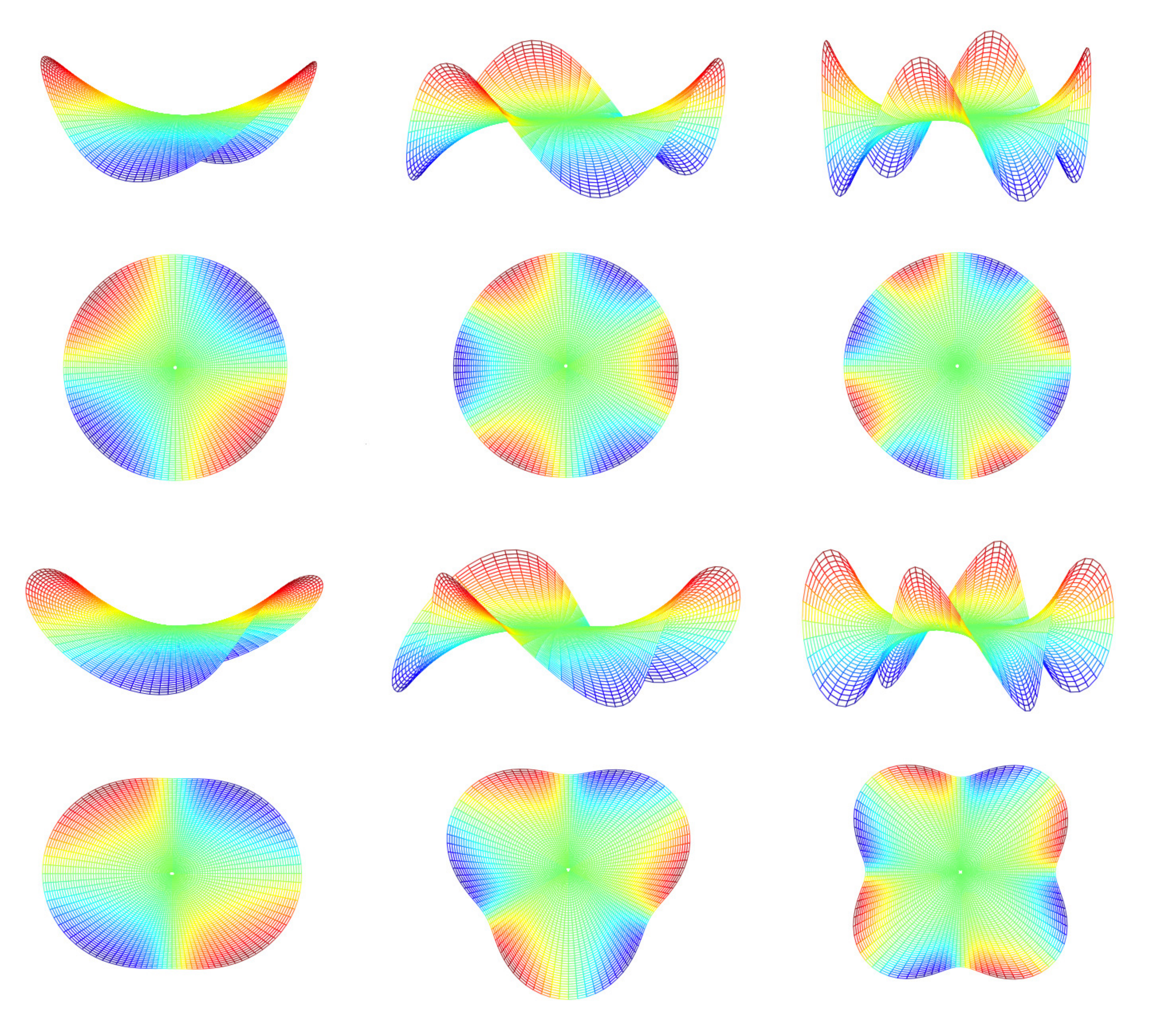}
\put(-342,355){\large$n=2$}
\put(-212,355){\large$n=3$}
\put(-82,355){\large$n=4$}
\caption{The bifurcation modes for modes $n=2$ to $4$. The linearized solutions in the first row correspond to \eqref{lin22} with the positive sign, while the solutions in the third row correspond to the negatively-signed alternative. The second and fourth rows depict projections of these surfaces onto the $(r,\theta)$-plane.}
\label{f2}
\end{figure}
With \eqref{lin19}, the kernels of the coefficient matrices appearing in \eqref{lin16} and \eqref{lin15} can be found.  On restricting attention to modes $n\ge2$, this determines the coefficients in \eqref{lin14} and \eqref{lin15}, and upon substituting in \eqref{decomp} and introducing
\begin{multline}
\bfeta_\pm(1,\theta)=-\frac{n+1}{2n\mu}\bigg(1\pm\sqrt{1+\frac{4n\mu^2}{n+1}}\,\bigg)(a_n \sin n\theta-b_n \cos n\theta)\er
\\[4pt]
+\frac{n+1}{2\mu n^2}\bigg(1\pm\sqrt{1+\frac{4n\mu^2}{n+1}}\,\bigg)(a_n \cos n\theta+b_n \sin n\theta)\et 
\\[4pt]
+(a_n \cos n\theta +b_n \sin n\theta)\ez,
\label{lin23}
\end{multline} 
leads to representations,
\be
\bfeta(1,\theta)=\left\{
\begin{array}{l l}
\displaystyle
\bfeta_\pm(1,\theta),   &\quad  \text{if} \quad \mu^2+1<n^2,
\cr\noalign{\vskip6pt}\displaystyle
\bfeta_+(1,\theta), & \quad \text{otherwise},
\end{array}
\right.
 \label{lin22}
 \ee
for the restriction to $r=1$ of the {in-plane} perturburation $\bfeta$. 

Substituting \eqref{lin22} in \eqref{perturbations}$_1$ determines the bounding curve of the surface governed by \eqref{lin5} and thereby delivers both the linear solution to the Euler--Lagrange equations and the fundamental modes that bifurcate from the disk-shaped ground state. Notice that the coefficients $a_n$ and $b_n$ in \eqref{lin23} are modal amplitudes, only one of which is independent when a  normalization condition on the perturbation field $\eta$ is imposed. Also, for $1+\mu^2<n^2$, the choice of sign in the first two terms on the right-hand side of \eqref{lin22} leads to two distinguished families of solutions. Representative plots of elements of these families are provided in Figure~\ref{f2}.  The projections of the first group on the $(r,\theta)$-plane are circles, and they reveal the $n$-fold rotational symmetry of mode $n$ around the {vertical} axis. {While the} primary bifurcation mode ($n=2$) is a simple saddle, the next one is a monkey saddle involving three depressions, and, similarly, the mode corresponding to $n=4$ is a saddle-like surface with four depressions. {Although members of the second family of solutions do not possess rotational symmetry, they are centrosymmetric.} In addition, their projections onto the $(r,\theta)$-plane exhibit $n$-fold symmetry. The projection of the primary bifurcation mode $n=2$ onto the the $(r,\theta)$-plane is ellipse-like and the subsequent modes yield $n$-fold star-shaped figures.  

\section{Discussion}
\label{discussion}
Figure~\ref{f3} compares the dimensionless parameters for the primary bifurcating mode $n=2$ in the present setting with those of the Euler--Plateau and Michell problems. In the absence of twisting energy, the curve on the right specializes to the result obtained for out-of-plane bifurcations in the context of the Euler--Plateau problem by Chen \& Fried~\cite{CF}. Also, the associated bifurcation mode, as depicted in the first row and column of Figure~\ref{f3}, agrees with their suggested saddle-shaped bifurcation mode. Although including twist does not alter the bifurcation mode, it increases the critical value of $\nu$ for transverse buckling---as might have been anticipated on intuitive grounds. Transverse buckling induces torsion on the boundary. Endowing the bounding loop with resistance to twisting effectively penalizes warping of the spanning surface. {Buckling of this kind therefore occurs only if the value of $\nu$ exceeds that predicted without twist.} 

On the other hand, if twist is neglected the curve on the left in Figure~\ref{f3} specializes to the planar elliptic bifurcation discussed by Chen \& Fried~\cite{CF}. Since a planar bifurcation is not accompanied by a change of the twisting energy, it is reasonable to ask why accounting for twist makes any difference and, moreover, why doing so leads to bifurcation at values of $\nu$ smaller than those arising in the Euler--Plateau problem. An answer to this question emerges on tracing the curve on the left in Figure~\ref{f3} down to $\nu=0$, where the value of $\mu$ corresponding to  Michell's \cite{Michell} instability is recovered. 
This indicates that the curve under consideration describes the bifurcations of a circular flexible loop which is twisted before joining its ends and is spanned by a soap film. To reduce its twisting energy, such a twisted loop buckles out of plane at $\mu=\sqrt{3}$. Any such bucking is, of course, penalized by an increase in the bending energy. If such a twisted ring is spanned by a soap film, the areal energy associated with the surface tension of the soap film also favors buckling. Under these circumstances, a less twisted circular loop therefore buckles out of plane. In other words, for a pre-twisted flexible loop, a flat, circular ground state becomes unstable at a smaller value of $\nu$. The bifurcation modes for this situation are depicted in the third row of Figure~\ref{f2}. These share characteristics with both the planar modes (ellipse-like projections onto the plane for the primary bifurcating mode) of the Euler--Plateau problem and the non-planar modes of Michell (see, for example, Goriely \& Tabor~\cite[p.~37]{GT1}). In the absence of twist ($\mu=0$), these bifurcation modes lose their nonplanar character and become similar to those depicted in the fourth row of Figure~\ref{f2}.

\section{Summary}
\label{conclusion}

The equilibrium of a flexible loop spanned by a soap film was investigated, accounting for the previously neglected effect of twist. The approach taken was inspired by various developments in Cosserat rod theory and by the parameterized formulation employed by pioneers (Alt~\cite{Alt}; Hildebrandt~\cite{Hild, LHD}; Nitsche~\cite{Nitsche}) in the study of the thread problem. The net potential-energy of the system, comprised by the bending and twisting energies of the flexible loop and the surface energy of the soap film, was varied under the constraint of lineal inextensibility to obtain the Euler--Lagrange equations. A flat, circular configuration was found to provide a trivial ground state solution to these equations. Small perturbations to the ground state were considered, leading to a linearized version of the Euler--Lagrange equations. In this linearized setting, the equilibrium of the surface is determined by the Laplace equation for the transverse component of the perturbation field. Moreover, the transverse and {in-plane} perturbations are coupled by a system of linear {boundary conditions}. 

Solutions to the linearized equilibrium equations were found to provide the basis for a linearized bifurcation analysis. A pair of dimensionless parameters were shown to control bifurcations from the ground state. While one of these is familiar from the Euler--Plateau problem, and measures the strength of the areal free-energy of the soap film relative to the lineal bending-energy of the flexible loop, the other combines the ratio of the twisting rigidity to the bending rigidity with a dimensionless measure of twist. In addition, two sets of bifurcation modes were identified. One set of bifurcation modes describes the buckling of a twisted {flexible loop} spanned by a soap film, which---if compared to the situation where the soap film is absent---loses its stability for smaller values of twist. For zero twist, these modes specialize to planar bifurcations of the Euler--Plateau problem. The second set of bifurcation modes, which describes transverse bifurcations, demonstrates the stabilizing role of twisting rigidity with respect to transverse buckling. 

{Aside from} its inherent physical and mathematical beauty, the problem considered here provides a simple model for various systems in which biological membranes are bonded by elastic filaments {or tubes}, such as high density lipoproteins, biofilms, and the dorsal mesentery. A natural generalization of {this} problem would therefore involve endowing the spanning surface with elastic resistance to bending. Under such conditions, the surface exerts not only a force but also a bending moment on the bounding loop (Biria, Maleki \& Fried~\cite{BMF}). Other future steps include studying stability and bifurcations from nontrivial equilibrium configurations, and solving the full nonlinear version of the problem---where the treatment of self-contact becomes inescapably important. The interplay between the elastic energy of intestinal tube and surface energy of the mesentery attached to it results in the coiled structure of the gut (Savin et al.~\cite{gut}), a property essential for fitting its length into the limited space of {abdominal} cavity. Nevertheless, excessive bending and twisting of the intestines, followed by knotting or kinking, result in volvulus and medical emergencies associated with bowel obstruction and ischemia. Solving the nonlinear version of Euler--Plateau with twist has the potential to shed light on health problems {stemming from} highly twisted intestines. 

Applications of the current problem are perhaps not limited to biological systems. For example, the boundary conditions of the Euler--Plateau problem have mathematical analogies in cosmology reminiscent of those considered by Eardley~\cite{BH}. Is this also true for the generalization of the Euler--Plateau problem considered here? Are there any similarities between shape of the bounding loop in the Euler--Plateau--Michell problem and the mysterious twisted ring of molecular cloud surrounding the region of hot gas and dust in the galactic center observed by Molinari et al.~\cite{cosmo}? These and other interesting questions remain open.

\newpage

%
%

\end{document}